\documentclass[prb,showpacs,twocolumn]{revtex4}
\usepackage{color,epsfig}
\usepackage{bm}
\usepackage{amsmath,amsfonts,amssymb}

\begin{document}
\title{Phonon-limited transport coefficients in extrinsic graphene}
\author{Enrique Mu\~noz}
\affiliation{
Instituto de F\'isica, Pontificia Universidad Cat\'olica de Valpara\'iso\\
PO Box 4059, Valpara\'iso, Chile.}

%\address{Instituto de F\'isica, Pontificia Universidad Cat\'olica de Valpara\'iso\\
%PO Box 4059, Valpara\'iso, Chile.}
\date{\today}

\begin{abstract}
The effect of electron-phonon scattering processes over the thermoelectric
properties of extrinsic graphene was studied. Electrical and thermal resistivity,
as well
as the thermopower, were calculated within the Bloch theory approximations.
Analytical expressions for the different transport coefficients
were obtained from a variational solution of the Boltzmann equation. The 
phonon-limited
electrical
resistivity $\rho_{e-ph}$
shows a linear dependence at high temperatures, and follows $\rho_{e-ph}\sim T^{4}$
at low temperatures, in agreement with experiments and theory previously reported in
the literature.
The phonon-limited thermal resistivity at low temperatures exhibits a $\sim T$
dependence, and
achieves a nearly constant value at high temperatures. The predicted Seebeck
coefficient at very low temperatures
is $Q(T) \sim -\pi^{2}k_{B}^{2}T/(3\,e\,E_{F})$, which shows
a $~ n^{-1/2}$ dependence with the density of carriers,
in agreement with experimental evidence. Our results suggest that thermoelectric properties
can be controlled by adjusting the Bloch-Gr\"uneisen temperature through its
dependence on the extrinsic carrier density in graphene.
\end{abstract}

\maketitle
Recently, Efetov and Kim \cite{Efetov.010} reported experimental measurements of the
phonon-limited electrical resistivity of graphene samples, by achieving extremely high carrier densities by means of an electrolytic gate, in order
to minimize the effect of other scattering mechanisms and to
be away from the neutrality (Dirac) point.
They observed that, at low temperatures, the
electrical resistivity displays a $\sim T^{4}$ dependence, whereas at higher
temperatures it follows a linear $\sim T$ trend, in agreement with an earlier theory by Hwang and Das Sarma \cite{Hwang.08}. The later based their analysis on the Bloch theory \cite{Ziman.60,Madelung.78}, which neglects phonon drag effects and umklapp processes, and modeled
the interaction between electrons and longitudinal acoustic phonons by the
deformation potential approximation. Moreover, they calculated the electrical
resistivity \cite{Hwang.08} within the relaxation time approximation, which is valid for elastic scattering processes, a condition not strictly satisfied
in electron-phonon scattering.

In this paper, phonon-limited 
transport coefficients in graphene are studied, by considering
the presence of temperature and voltage gradients. Under similar
assumptions as in Ref. (\cite{Hwang.08}), that is, Bloch theory and deformation potential
approximation for the interaction between electrons and longitudinal-acoustic phonons,
analytical expressions for the transport coefficients
are obtained from a variational solution of the Boltzmann equation \cite{Ziman.60, Madelung.78}. This method, at
the lowest-order approximation, provides equivalent results to the relaxation time
approximation and, at higher orders, converges to solutions of the Boltzmann
equation which are exact to an arbitrary precision, within
the limitations of the set of variational functions chosen.
As has been pointed out before \cite{Peres.07}, the semiclassical approach
based on the Boltzmann equation cannot capture the subtleties of electronic transport in graphene right at the neutrality point. This limitation can, at least in part, be attributed to the inhomogeneity
in the charge distribution experimentally observed in graphene samples at the neutrality point\cite{Peres.010}. However, at high carrier densities the Fermi level
is well above the Dirac point, and the Boltzmann equation then provides
a correct description.

\begin{figure}[tbp]
\centering
\epsfig{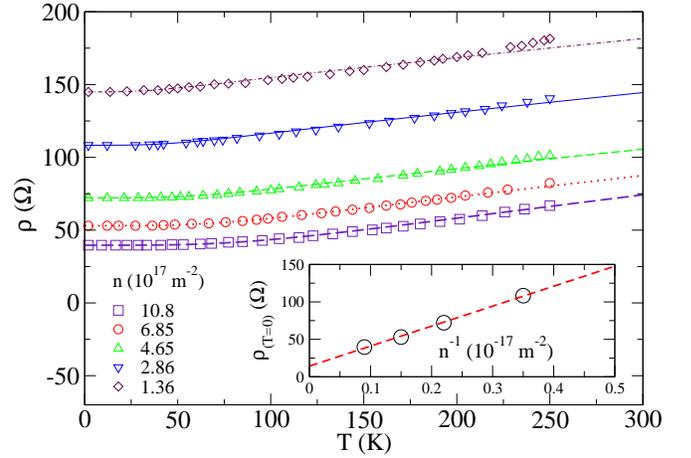}
\caption{(Color online) Electrical resistivity calculated from Eq. (\ref{eq33}) (lines), compared with experimental data (symbols) reported in Ref.(\cite{Efetov.010}) at different carrier densities $n$ in units
of $10^{17} m^{-2}$. Inset: Electrical resistivity at zero temperature, from
data in Ref. (\cite{Efetov.010}) (circles), at different carrier densities $n$. Also shown is the linear fit (dashed line) $\rho_{(T=0)} = 14.276 + 266.17\times10^{17}\,n^{-1}$ ($\Omega$) explained in the main text.
\label{fig1}
}
\end{figure}
\section{Electron-phonon interaction}

As is well known from tight-binding calculations, electronic transport
in graphene displays relativistic features. This is because pristine
graphene, free of impurities and defects, behaves as a metal only in
the so-called Dirac points where the conduction and valence bands touch. 
There exists two non-equivalent such points within the first Brillouin zone, defined by the vectors $\mathbf{K}_{1,2}=\pm\frac{2\pi}{a}\left[1/\sqrt{3}\,,\,1\right]$, with $a$ the direct lattice parameter. 
In the vicinity of each of these points, or valleys, it is shown that the dispersion
relation is linear, $E_{\eta}(\delta\mathbf{k}) = \pm\hbar v_{F} |\delta\mathbf{k}|$,
for $\delta\mathbf{k} = \mathbf{k} - \mathbf{K}_{\eta}$. 
In each valley $\mathbf{K}_{\eta}$, the envelope
electronic functions related to this dispersion relation
can thus be obtained as the pseudo-spinor solutions of an effective two-dimensional Dirac's equation, 
\begin{eqnarray}
\hat{H}_{\mathbf{K}}\psi_{\mathbf{k}}^{(\pm)} = v_{F}\hat{\mathbf{\alpha}}\cdot\hat{\mathbf{p}}\psi_{\mathbf{k}}^{(\pm)} = E_{k}^{(\pm)}\psi_{\mathbf{k}}^{(\pm)}.
\label{eq1}
\end{eqnarray}
Here $\hat{\mathbf{\alpha}}=(\hat{\sigma}_{x},\hat{\sigma}_{y})$
are the Pauli matrices, we have re-defined $\mathbf{k}\equiv\delta\mathbf{k}$ for each valley, and the energy eigenvalues for positive (negative) helicity are $E_{k}^{(\pm)} = \pm\hbar v_{F}k$. The corresponding
eigenvectors, normalized by the total area $A$ of the sample, are given by the pseudo-spinors
\begin{eqnarray}
\psi_{\mathbf{k}}^{(\pm)}(\mathbf{r}) = \frac{1}{\sqrt{2A}}\left( \begin{array}{c} e^{-i\varphi_{k}/2}\\ \pm e^{i\varphi_{k}/2}\end{array}\right)e^{i\mathbf{k}\cdot\mathbf{r}}\equiv u_{\mathbf{k}}^{(\pm)}\frac{e^{i\mathbf{k}\cdot\mathbf{r}}}{\sqrt{A}}
\label{eq2}
\end{eqnarray}
with $\varphi_{k} = \arctan(k_{y}/k_{x})$.

In the language of second quantization, the conduction electrons field for 
$\mathbf{k}$ in the vicinity of
the $\mathbf{K}_{1,2}=\pm\frac{2\pi}{a}\left[1/\sqrt{3}\,,\,1\right]$ valley is
\begin{eqnarray}
\hat{\psi}_{\eta}(\mathbf{r}) = \sum_{\mathbf{k},\sigma}\psi_{\mathbf{k}}^{(+)}(\mathbf{r})\hat{c}_{\mathbf{k}\sigma\eta}
\label{eq3}
\end{eqnarray}
with  $\hat{c}_{\mathbf{k}\sigma\eta}$ Fermionic operators $[\hat{c}_{\mathbf{k}\sigma\eta},\hat{c}_{\mathbf{k}'\sigma'\eta'}^{\dagger}]_{+} = \delta_{\mathbf{k},\mathbf{k}'}\delta_{\sigma,\sigma'}\delta_{\eta,\eta'}$,  that destroy (create) a conduction
electron with momentum $\hbar\mathbf{k}$ in the vicinity of the
$\mathbf{K}_\eta$ valley, with spin component $\sigma = \{\uparrow,\downarrow\}$.
We model the electron-phonon interaction by the deformation potential approximation \cite{Hwang.08}, which is a reasonable one when
the Fermi surface possesses spherical symmetry \cite{Ziman.60,Madelung.78}. The two-dimensional version of this is a Fermi "circle" in graphene, which
corresponds to the intersection of the Dirac cone and the constant energy plane defined by the Fermi energy at $\epsilon_{k} = E_{F}$.
In the deformation potential approximation, the operator representing lattice dilation in the second quantization
language is
\begin{eqnarray}
\hat{\Delta}(\mathbf{r}) = i\sum_{\mathbf{q}}\left(2 A \rho_{m} \omega_{q} \right)^{-1/2}(\mathbf{q}\cdot\hat{e}_{L})\left(
\hat{a}_{q}^{\dagger}e^{i\mathbf{q}\cdot\mathbf{r}} - \hat{a}_{q}e^{-i\mathbf{q}\cdot\mathbf{r}} \right) 
\label{eq4}
\end{eqnarray}
Here, $\omega_{q} = v_{s}q$ is the frequency for longitudinal acoustic phonons in graphene, $\hat{a}^{\dagger}_{q}$ is the creation operator
for longitudinal acoustic phonon modes, and $\hat{e}_{L}$ is the corresponding polarization vector for these modes.
The electron-phonon interaction Hamiltonian, in the deformation potential approximation, is obtained from
\begin{eqnarray}
&&\hat{H}^{int}_{e-ph} = i\int {d^{2}r}\sum_{\eta=1,2}\hat{\psi}^{\dagger}_{\eta}(\mathbf{r})D\hat{\Delta}(\mathbf{r})\hat{\psi}_{\eta}(\mathbf{r})\\
&&= i\sum_{\mathbf{k},\mathbf{q},\eta,\sigma}\sqrt{\frac{\hbar q^{2}D^{2}}{2\rho_{m}\omega_{q}}}\cos\left(\theta/2\right)
\left(\hat{a}_{\mathbf{q}} - \hat{a}_{-\mathbf{q}}^{\dagger}
\right)\hat{c}^{\dagger}_{\mathbf{k}+\mathbf{q},\sigma\eta}\hat{c}_{\mathbf{k}\sigma\eta}\nonumber
\label{eq5}
\end{eqnarray}
Notice the presence of the factor $\cos(\theta/2) = \cos([\varphi_{k}-\varphi_{k'}]/2)$ which arises from the inner product
of the two pseudo-spinor functions $u_{\mathbf{k}'}^{*}\cdot u_{\mathbf{k}}$ defined in Eq. (\ref{eq2}). This factor, which would be absent from the inner product of non-relativistic scalar Schr\"odinger
wave-functions, is a fingerprint of the relativistic features of
electrons in graphene. The more obvious consequence of its presence the suppression of
backscattering ($\theta = \pi$), a phenomenon observed
in graphene and directly related to so called Klein tunneling \cite{Beenakker.08,Peres.010,Castro.09} in the context of relativistic quantum mechanics.

We shall
neglect umklapp processes, and thus we have
 $\mathbf{G}=0$ for normal processes. In
this case, scattering events satisfy energy and "crystal momentum" conservation,
$\epsilon_{k'} = \epsilon_{k} \pm \hbar\omega_{q}$ and
$\mathbf{k}' = \mathbf{k} \pm \mathbf{q}$. From the latter equation, one obtains the
condition $q^{2} = k'^{2} + k^{2} - 2k k'
cos(\theta)$, with $\theta$ the scattering angle. Since the
relevant scattering events occur within a thin layer
near the Fermi surface,  then $k \sim k' \sim k_{F}$, which then
implies $\left|sin(\theta/2)\right| = q/2k_{F}$. In a Debye
model approximation,
this condition restricts the range of lattice momenta involved in electron-phonon
scattering,
$q \le \min\{2 k_{F}, q_{D} \}$, with $q_{D}$ the radius of the Debye circle
(for a two-dimensional system). In geometrical terms, the most restrictive condition
depends on whether the radius of the Fermi surface (a circle centered on each Dirac
point for graphene) is larger or smaller than the Debye circle. When written in terms of
the frequency of the longitudinal-acoustic phonon modes involved $\omega_{q} =
v_{s}q$, the former condition reads $\hbar \omega_{q}/k_{B} \le
\min\{\Theta_{BG},\Theta_{D}\}$, with $\Theta_{D}$ the Debye temperature, and
$\Theta_{BG} = 2(v_{s}/v_{F})E_{F}/k_{B}$ the Bloch-Gr\"uneisen temperature. In the
experimental setup presented in Ref.(\cite{Efetov.010}), the carrier densities involved are such
that the Fermi surface is contained within the Debye circle, and therefore the
Bloch-Gr\"uneisen rather than the Debye temperature imposes the characteristic
energy scale for phonon scatterers. Notice that the analysis of the scattering
angle above also yields the relation $\cos(\theta/2) = \sqrt{1 - (q/(2k_{F}))^{2}}$.
Therefore, after Eq. (\ref{eq5})
the matrix element for
the electron-phonon interaction $\hat{H}_{e-ph}^{int}=i\sum_{\mathbf{k},\mathbf{q},\eta,\sigma}M_{\mathbf{q}}\left(\hat{a}_{\mathbf{q}} - \hat{a}_{-\mathbf{q}}^{\dagger}
\right)\hat{c}^{\dagger}_{\mathbf{k}+\mathbf{q},\sigma\eta}\hat{c}_{\mathbf{k}\sigma\eta}$ is
\begin{eqnarray}
M_{q} = -i\left(\hbar/2\rho_{m}\omega_{q}\right)^{1/2}D q
\left(1-\left(q/2k_{F}\right)^{2}\right)^{1/2}.
\label{eq6}
\end{eqnarray}
Here, $D$  is the deformation potential\cite{Ziman.60,Madelung.78,Efetov.010} coupling constant for
graphene, and $\rho_{m}=7.6\times 10^{-7}$
kg/$\rm{m}^{2}$
the surface mass density.

\section{Boltzmann Equation}
We introduce, in the usual form, the deviation of the distribution function
from equilibrium by the expression $f_{k} = f_{k}^{0} - \chi_{\mathbf{k}}\partial f_{k}^{0}/\partial \epsilon_{k}$.
Let us consider the electron-phonon interaction as the sole scattering mechanism. Therefore,
after Eqs.(5,\ref{eq6}) we should consider processes of the form $\mathbf{k} + \mathbf{q} \rightleftarrows \mathbf{k}'$. Under Bloch's
approximation, which is to assume that the phonon system is in quasi-equilibrium, the linearized form of the Boltzmann equation when both an external electric field
and a temperature gradient are imposed can be expressed as
\begin{eqnarray}
&&-\mathbf{v}_{\mathbf{k}}\cdot \nabla T \frac{\partial f_{k}^{0}}{\partial T} - \mathbf{v}_{\mathbf{k}}\cdot\mathbf{E}(-e)\frac{\partial f_{k}^{0}}{\partial \epsilon_{k}}
= \nonumber\\
&&\frac{g_{\eta}g_{\sigma}}{k_{B}T}\sum_{\mathbf{k}',\mathbf{q}}\left[\{\chi_{\mathbf{k}} - \chi_{\mathbf{k}'} \}\mathcal{P}_{\mathbf{k}\mathbf{q}}^{\mathbf{k}'}
-\{\chi_{\mathbf{k}'} - \chi_{\mathbf{k}} \}\mathcal{P}^{\mathbf{k}\mathbf{q}}_{\mathbf{k}'}\right]
\label{eq7}
\end{eqnarray}
Here, $g_{\eta}=g_{\sigma}=2$ are the spin and valley degeneracies, respectively, while the transition rate
for electron-phonon scattering processes is
\begin{eqnarray}
\mathcal{P}_{\mathbf{k}\mathbf{q}}^{\mathbf{k}'} &&= \left(2\pi/\hbar\right)
\delta_{\mathbf{G},\mathbf{k}' - \mathbf{k} - \mathbf{q}}
|M_{q}|^{2}\delta(\epsilon_{k}-\epsilon_{k'} + \hbar\omega_{q})\nonumber\\
&&\times n_{q}^{0}f_{k}^{0}(1 - f^{0}_{k'})
\label{eq8}
\end{eqnarray}
Here  $f_{k}^{0}$ is the
Fermi-Dirac distribution for electron states with momentum $\mathbf{k}$,
and $n_{q}^{0}$ the Bose distribution for phonon states with momentum $\mathbf{q}$. All vectors are two-dimensional, with $\mathbf{G}$ a vector of the reciprocal lattice.

For extrinsic graphene, the Fermi level $E_{F} = \hbar k_{F} v_{F}$ is located above
the Dirac point, and depends on the density of carriers $n$ by the relation
$k_{F} = \sqrt{\pi n}$. In Ref.(\cite{Efetov.010}), an experimental method is presented which
allows to control the carrier density, and hence the Bloch-Gr\"uneisen temperature. As
discussed in Refs.(\cite{Efetov.010, Fuhrer.010}), this in turn becomes a way to control the phonon-limited
electrical resistivity. In this work, we will show that also
the electronic component of the thermal conductivity, as well as the thermopower
(Seebeck coefficient) which determines thermoelectric effects, can be controlled in a similar way. We shall use the
variational method to calculate the phonon-limited transport coefficients
which, in contrast with the relaxation
time approximation, it has the advantage of avoiding the assumption of quasi-elasticity
in the scattering processes.
\begin{figure}[tbp]
\centering
\epsfig{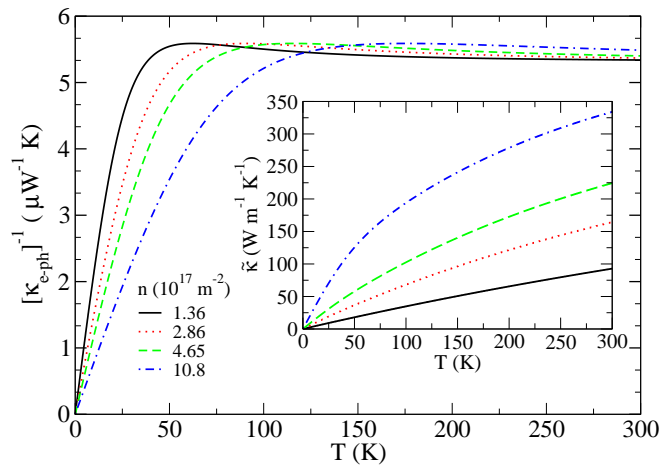}
\caption{(Color online) Phonon-limited thermal resistivity $[\kappa_{e-ph}]^{-1}$, calculated from Eq.(\ref{eq37}) at different carrier densities $n$, in units
of $10^{17} m^{-2}$. Shown in the inset is the corresponding thermal conductivity, including the contribution from Coulomb elastic scattering with
impurities, and normalized with a nominal packing thickness of $3.4$ \AA.
\label{fig2}
}
\end{figure}

\section{Coupled charge and heat transport}

For a system with thermal and electrostatic potential gradients, the
macroscopic entropy production rate is given in terms of the heat $\mathbf{U}$ and charge
$\mathbf{J}$ fluxes by \cite{Ziman.60,Madelung.78}
\begin{eqnarray}
\partial_{t}S = \mathbf{J}\cdot\mathbf{E}/T +
\mathbf{U}\cdot\nabla\left(1/T\right).
\label{eq9}
\end{eqnarray}
The Onsager's theorem \cite{Onsager.31} indicates that the fluxes and generalized potentials
are then linearly coupled,
\begin{eqnarray}
\mathbf{J} = L_{EE}\mathbf{E} + L_{ET}\nabla T \nonumber\\
\mathbf{U} = L_{TE}\mathbf{E} + L_{TT}\nabla T
\label{eq10}
\end{eqnarray}
The thermopower $Q$ can be expressed in terms of the coefficients $L_{\alpha\beta}$
by considering
the case $\mathbf{J}=0$ in Eq. (\ref{eq10}), to obtain
$\mathbf{E} = -L_{ET}L_{EE}^{-1}\nabla T$. Hence, one concludes that $Q =
-L_{ET}L_{EE}^{-1}$. The thermal conductivity is obtained
from the second equation, $\mathbf{U} = -(L_{TE}L_{EE}^{-1}L_{ET}-L_{TT})\nabla T$,
which implies $\kappa = L_{TE}L_{EE}^{-1}L_{ET}-L_{TT}$. If, on the other hand, one
considers the case $\nabla T = 0$, then the relation $\mathbf{J}=L_{EE}\mathbf{E}$ is
obtained. Thus, it is concluded that the electrical resistivity is $\rho = L_{EE}^{-1}$. The variational method \cite{Ziman.60,Madelung.78} is based on the principle of maximal entropy
production \cite{Onsager.31}, and
provides an iterative but virtually exact procedure to solve the Boltzmann equation,
and therefore to calculate the transport coefficients, within the limitations of the
set of variational functions chosen. The principle is based on equating the macroscopic
entropy production rate, as expressed in terms of the macroscopic currents Eq.(\ref{eq10}), with
the entropy production rate due to microscopic scattering events\cite{Onsager.31}, $\partial_{t}S = \dot{S}_{scatt}$,
as defined from the linearized Boltzmann equation scattering terms,
\begin{eqnarray}
\dot{S}_{scatt} = \frac{g_{\eta}g_{\sigma}}{k_{B}T^{2}}\sum_{\mathbf{k},\mathbf{q},\mathbf{k}'}\{\chi_{\mathbf{k}} - \chi_{\mathbf{k}'} \}^{2}\mathcal{P}_{\mathbf{k}\mathbf{q}}^{\mathbf{k}'}
\label{eq11}
\end{eqnarray}
The electric current $\mathbf{J}$ and the heat current $\mathbf{U}$ are given by the expressions
\begin{eqnarray}
\mathbf{J} &=& g_{\eta}g_{\sigma}\sum_{\mathbf{k}}(-e)\mathbf{v}_{\mathbf{k}}\chi_{\mathbf{k}}\frac{\partial f_{k}^{0}}{\partial \epsilon_{k}}\nonumber\\
\mathbf{U} &=& g_{\eta}g_{\sigma}\sum_{\mathbf{k}}\mathbf{v}_{\mathbf{k}}(\epsilon_{k}-E_{F})\chi_{\mathbf{k}}\frac{\partial f_{k}^{0}}{\partial \epsilon_{k}}
\label{eq12}
\end{eqnarray}
The technique is based on expanding the distribution function deviation
from equilibrium $\chi_{\mathbf{k}} = \sum_{i}\alpha_{i}\phi_{i}(\mathbf{k})$ in terms of a set of variational functions of the form $\phi_{i}(\mathbf{k}) = (\epsilon_{k} - E_{F})^{i-1}(\hat{u}\cdot\mathbf{k})$.
For a minimal set of trial functions
$\phi_{1}(\mathbf{k})=\mathbf{k}\cdot\hat{u}$ and
$\phi_{2}(\mathbf{k})=\mathbf{k}\cdot\hat{u}(\epsilon_{k}-E_{F})$, we
obtain the charge and heat currents
\begin{eqnarray}
\mathbf{J}_{i} &&= (-e) g_{\eta}g_{\sigma}\sum_{\mathbf{k}}
\mathbf{v}_{\mathbf{k}}\phi_{i}(\mathbf{k})\frac{\partial{f}^{0}}{\partial\epsilon_{k}}\nonumber\\
\mathbf{U}_{i} &&= g_{\eta}g_{\sigma}\sum_{\mathbf{k}} \mathbf{v}_{\mathbf{k}}(\epsilon_{k} -
E_{F})\phi_{i}(\mathbf{k})\frac{\partial{f}^{0}}{\partial\epsilon_{k}}
\label{eq13}
\end{eqnarray}
Here, $g_{\eta}=g_{\sigma}=2$ are the valley and spin degeneracies in graphene, respectively.
The coefficients $P_{ij} = P_{ji}$ are defined as follows
\begin{eqnarray}
P_{ij} = \frac{(g_{\eta}g_{\sigma})^{2}}{k_{B}T}\sum_{\mathbf{k},\mathbf{k}',\mathbf{q}}\{\phi_{i}(\mathbf{k})-\phi_{i}(\mathbf{k}')\}
\{\phi_{j}(\mathbf{k})-\phi_{j}(\mathbf{k}')\}\mathcal{P}_{\mathbf{k}\mathbf{q}}^{\mathbf{k}'}\nonumber\\
\label{eq14}
\end{eqnarray}
In terms of the variational solution, after Eq.(\ref{eq12}) the electric and heat currents
are written as
\begin{eqnarray}
\mathbf{J} &=& \sum_{i}\alpha_{i}\mathbf{J}_{i}\nonumber\\
\mathbf{U} &=& \sum_{i}\alpha_{i}\mathbf{U}_{i}
\label{eq15}
\end{eqnarray}
whereas the macroscopic $\partial_{t}S$ and microscopic $\dot{S}_{scatt}$
entropy generation rates in Eqs.(\ref{eq9},\ref{eq11}) are expressed by
\begin{eqnarray}
\partial_{t}S &=& \sum_{i}\alpha_{i}\left[T^{-1}\mathbf{J}_{i}\cdot\mathbf{E}
- T^{-2} \mathbf{U}_{i}\cdot\mathbf{\nabla}T \right],
\nonumber\\
\dot{S}_{scatt} &=& T^{-1}\sum_{i,j}\alpha_{i}\alpha_{j}P_{ij}.
\label{eq16}
\end{eqnarray}
The variational problem is to find the set of coefficients $\{\alpha_{j} \}$ that maximize $\dot{S}_{scatt}$, subject to the constraint 
$\partial_{t}S = \dot{S}_{scatt}$, which is enforced by a Lagrange multiplier $\lambda$,
\begin{eqnarray}
\frac{\delta}{\delta\alpha_{n}}\left[\sum_{i,j}\alpha_{i}\alpha_{j}\frac{P_{ij}}{T} - \lambda\,\,\sum_{i}\alpha_{i}\left(\frac{\mathbf{J}_{i}\cdot\mathbf{E}}{T}- \mathbf{U}_{i}\cdot\frac{\mathbf{\nabla}T}{T^{2}}  \right)\right] = 0\nonumber\\
\label{eq17}
\end{eqnarray}
The solution to this constrained variational problem is given by $\lambda = 1/2$, and
\begin{eqnarray}
\alpha_{i} = \sum_{j}[P^{-1}]_{ij}\left(\mathbf{J}_{j}\cdot\mathbf{E} - \mathbf{U}_{j}\cdot\frac{\nabla T}{T} \right)
\label{eq18}
\end{eqnarray}
Here, $[P^{-1}]$ is the inverse of the matrix $[P]_{ij} = P_{ij}$
whose elements are the coefficients $P_{ij}$ defined by Eq.(\ref{eq14}). After substituting this solution into the expressions for the electric and heat currents, we obtain explicit analytical formulas for the
transport coefficient tensors defined in Eq.(\ref{eq10}),
\begin{eqnarray}
L_{EE} &=& \sum_{ij}\mathbf{J}_{i}[P^{-1}]_{ij}\mathbf{J}_{j}\nonumber\\
L_{ET} &=& -T^{-1}\sum_{ij}\mathbf{J}_{i}[P^{-1}]_{ij}\mathbf{U}_{j}\nonumber\\
L_{TE} &=& \sum_{ij}\mathbf{U}_{i}[P^{-1}]_{ij}\mathbf{J}_{j}\nonumber\\
L_{TT} &=& -T^{-1}\sum_{ij}\mathbf{U}_{i}[P^{-1}]_{ij}\mathbf{U}_{j}
\label{eq19}
\end{eqnarray}
Therefore, we proceed to calculate the currents and $P_{ij}$ coefficients.
Let us first consider the current $\mathbf{J}_{1}$. 
The group velocity in Eq.(\ref{eq13}) is given by
$\mathbf{v}_{\mathbf{k}} =\hat{k}v_{F}$, with $\hat{k} = \mathbf{k}/k = (\cos(\varphi),\sin(\varphi))$.
The discrete sums are treated in the usual quasi-continuum limit as two-dimensional integrals, $\sum_{\mathbf{k}}\rightarrow A(2\pi)^{-2}\int d^{2}k$, and the integral is calculated by using the change of variables
$d^{2}k = d\varphi \epsilon d\epsilon/(\hbar v_{F})^{2}$,
\begin{eqnarray}
\mathbf{J}_{1}&&=\frac{A (-e) v_{F}}{\pi^{2}(\hbar v_{F})^{3}}\int_{-\pi}^{+\pi}d\varphi \hat{k}\hat{k}\cdot\hat{u}\int d\epsilon \epsilon^{2} \frac{\partial f_{\epsilon}^{0}}{\partial \epsilon} \nonumber\\
&&= \hat{u}\frac{A e}{\pi\hbar}\left(\frac{E_{F}}{\hbar v_{F}}\right)^{2}
\label{eq20}
\end{eqnarray}
For the current $\mathbf{U}_{1}$, we consider the quasi-continuum limit for the sum over wave-vectors in Eq.(\ref{eq13}), with directions
$\hat{k}=\mathbf{k}/k = (\cos(\varphi),\sin(\varphi))$. Using
the integration variables $d^{2}k = d\varphi d\epsilon \epsilon/(\hbar v_{F})^{2}$, we have
\begin{eqnarray}
\mathbf{U}_{1} &=& \frac{A v_{F}}{\pi^{2}(\hbar v_{F})^{3}}\int_{-\pi}^{\pi}d\varphi\hat{k}\hat{k}\cdot\hat{u}\int d\epsilon \epsilon^{2}(\epsilon - E_{F})\left(\frac{\partial f_{\epsilon}^{0}}{\partial \epsilon} \right)\nonumber\\
&=& -\hat{u}\frac{2\pi A}{3}\frac{(k_{B}T)^{2}}{\hbar}\frac{E_{F}}{(\hbar v_{F})^{2}}
\label{eq20b}
\end{eqnarray}
Similarly, for the current $\mathbf{U}_{2}$ we obtain
\begin{eqnarray}
\mathbf{U}_{2} &=& \frac{A v_{F}}{\pi^{2}(\hbar v_{F})^{3}}\int_{-\pi}^{\pi}d\varphi\hat{k}\hat{k}\cdot\hat{u}\int d\epsilon \epsilon^{2}(\epsilon - E_{F})^{2}\left(\frac{\partial f_{\epsilon}^{0}}{\partial \epsilon} \right)\nonumber\\
&=& -\hat{u}\frac{\pi A}{3}\frac{(k_{B}T)^{2}}{\hbar}\left(\frac{E_{F}}{\hbar v_{F}} \right)^{2}
\label{eq21}
\end{eqnarray}
Finally, it is straightforward to verify by inspection that $\mathbf{J}_{2}=(-e)\mathbf{U}_{1}$.

Let us now consider the coefficient $P_{11}$.
From Eq.(\ref{eq8}), we first notice that the transition probability rate 
$\mathcal{P}_{\mathbf{k}\mathbf{q}}^{\mathbf{k}'}$ imposes the condition $\mathbf{q} =\mathbf{k}' - \mathbf{k}$. We define the directions $\hat{k}'=(\cos(\varphi'),\sin(\varphi'))$
and $\hat{k}=(\cos(\varphi' - \theta),\sin(\varphi' - \theta))$, with
$\theta$ the scattering angle. As before, in the quasi-continuum limit for the sums over wave-vectors, we choose the integration variables $d^{2} k d^{2} k' = d\varphi'  d\theta d\epsilon' \epsilon' d\epsilon \epsilon/(\hbar v_{F})^{4}$,
to obtain
\begin{eqnarray}
P_{11} &&= \frac{(k_{B}T)^{-1}A^{2}}{(\pi \hbar v_{F})^{4}}
\int d\epsilon \epsilon \int d\epsilon' \epsilon'\nonumber\\
&&\times\int_{-\pi}^{\pi}d\theta\int_{-\pi}^{\pi}d\varphi' (\mathbf{q}\cdot\hat{u})^{2}\mathcal{P}_{\mathbf{k}\mathbf{q}}^{\mathbf{k}'}
\label{eq22}
\end{eqnarray}
In Appendix A it is shown that the angular integral
$\int d\varphi' (\mathbf{q}\cdot\hat{u})^{2} = \pi q^{2}$, with $q = 2k_{F}\sin(\theta/2)$. Hence, substituting the explicit expression for the transition probability rate, Eq.(\ref{eq22})
reduces to
\begin{eqnarray}
&&\frac{2(k_{B}T)^{-1}A^{2}}{\pi^{2} \hbar(\hbar v_{F})^{4}}\int_{-\pi}^{\pi}d\theta q^{2}n_{q}^{0}|M_{q}|^{2}\nonumber\\
&&\times\int d\epsilon
\epsilon (\epsilon + \hbar\omega_{q})f_{\epsilon}^{0}(1 - f_{\epsilon + \hbar\omega_{q}}^{0})
\label{eq23}
\end{eqnarray}
The energy integral is evaluated using the technique presented in Appendix B. Using the change of variables $d\theta = k_{F}^{-1}[1 - (q/(2k_{F}))^{2}]^{-1/2}dq$,
we obtain
\begin{eqnarray}
&&P_{11}=\frac{(k_{B}T)^{-1}A^{2}k_{F}\hbar D^{2}}{\pi^{2}\rho_{m}(\hbar v_{F})^{2}}\int_{0}^{2k_{F}}
dq \left(q^{4}+4\left(\frac{v_{s}}{v_{F}}\right)^{2}\right.\nonumber\\
&&\times\left.\left(\frac{T}{\Theta_{BG}} \right)^{2}\left[\frac{\pi^{2}q^{4}}{3}+\frac{q^{6}}{2} \right][1 - \left(\frac{q}{2k_{F}}\right)^{2}]^{1/2}\right)\nonumber\\
&&\times(1 - e^{-\hbar\omega_{q}/(k_{B}T)})^{-1}(e^{\hbar\omega_{q}/(k_{B}T)}-1)^{-1}
\label{eq24}
\end{eqnarray}
Eq.(\ref{eq24}) is written in terms of the functions $\mathcal{J}_{p}(z)$
defined in Appendix C,
\begin{eqnarray}
&&P_{11}=\frac{\hbar D^{2}A^{2}E_{F}(k_{B}T)^{4}}{\pi^{2}\rho_{m}(\hbar v_{s})^{5}(\hbar v_{F})^{3}}\left[\mathcal{J}_{4}(T/\Theta_{BG}) + 4\left(\frac{v_{s}}{v_{F}} \right)^{2}\right.\nonumber\\
&&\left.\times\left(\frac{T}{\Theta_{BG}} \right)^{2}\left\{\frac{\pi^{2}}{3}\mathcal{J}_{4}(T/\Theta_{BG})
 +\frac{1}{2}\mathcal{J}_{6}(T/\Theta_{BG})\right\} \right]
\label{eq25}
\end{eqnarray}
Let us now consider the coefficient $P_{22}$, which after similar manipulations as before, can be expressed by the integral form
\begin{eqnarray}
P_{22} &&= \frac{(k_{B}T)^{-1}A^{2}}{(\pi \hbar v_{F})^{4}}
\int_{-\pi}^{\pi}d\theta\int_{-\pi}^{\pi}d\varphi'\int d\epsilon \epsilon \int d\epsilon' \epsilon'\nonumber\\
&&\times [(\mathbf{q}\cdot\hat{u})(\epsilon-E_{F}) +  \hbar\omega_{q}(\mathbf{k}'\cdot\hat{u})]^{2}\mathcal{P}_{\mathbf{k}\mathbf{q}}^{\mathbf{k}'}
\label{eq26}
\end{eqnarray}
After expanding the square in Eq.(\ref{eq26}), it is shown in Appendix A that the angular integrals give,
$\int d\varphi' (\mathbf{q}\cdot\hat{u})^{2} = \pi q^{2}$, $\int d\varphi' (\mathbf{k}'\cdot\hat{u}) = \pi k'^{2}$, and $\int d\varphi'(\mathbf{q}\cdot\hat{u})(\mathbf{k}'\cdot\hat{u}) = (\pi/2) [q^{2} + k'^{2}-k^{2}]$.
Then, 
after similar algebraic procedures as described previously, Eq.(\ref{eq26}) reduces to
\begin{eqnarray}
&&P_{22}=\frac{\hbar D^{2}A^{2}E_{F}^{3}(k_{B}T)^{4}}{\pi^{2}\rho_{m}(\hbar v_{s})^{3}(\hbar v_{F})^{5}}
\left[
\left\{1 + \frac{4\pi^{2}}{3}
\left(1 + 12\right.\right.\right.\nonumber\\
&&\left.\left.\left.\times\left(\frac{v_{s}}{v_{F}} \right)^{2} \right) \right\}\left(\frac{T}{\Theta_{BG}} \right)^{2}\mathcal{J}_{4}(T/\Theta_{BG})-\frac{2}{3}\left\{1 - 6
\right.\right.\nonumber\\
&&\left.\left.
\times
\left(\frac{v_{s}}{v_{F}} \right)^{2} \right\}
\left(\frac{T}{\Theta_{BG}} \right)^{2}\mathcal{J}_{6}(T/\Theta_{BG}) + \frac{16\pi^{2}}{3}\left(\frac{v_{s}}{v_{F}}\right)^{2}
\right.\nonumber\\
&&\left.\times
\left(\frac{T}{\Theta_{BG}} \right)^{4}
\left\{\left(\frac{v_{s}}{v_{F}}\right)^{2}\mathcal{J}_{6}(T/\Theta_{BG})
+\frac{1}{10\pi^{2}}\left\{1\right.\right.\right.\nonumber\\
&&\left.\left.\left.
-2\left(\frac{v_{s}}{v_{F}}\right)^{2}\right\}
\mathcal{J}_{8}(T/\Theta_{BG})
\right\}
\right]
\label{eq28}
\end{eqnarray}
At last, we consider the coefficient $P_{12}$. After our definition Eq.(\ref{eq14}), we have
\begin{eqnarray}
P_{12} &&= \frac{(k_{B}T)^{-1}A^{2}}{(\pi \hbar v_{F})^{4}}
\int_{-\pi}^{\pi}d\theta\int_{-\pi}^{\pi}d\varphi'\int d\epsilon \epsilon \int d\epsilon' \epsilon'\nonumber\\
&&\times [(\mathbf{q}\cdot\hat{u})^{2}(\epsilon-E_{F}) + \hbar\omega_{q}(\mathbf{k}'\cdot\hat{u})(\mathbf{q}\cdot\hat{u})]\mathcal{P}_{\mathbf{k}\mathbf{q}}^{\mathbf{k}'}
\label{eq29}
\end{eqnarray}
As shown in Appendix A, the angular integrals are given by
$\int d\varphi' (\mathbf{q}\cdot\hat{u})^{2} = \pi q^{2}$, and $\int d\varphi'(\mathbf{q}\cdot\hat{u})(\mathbf{k}'\cdot\hat{u}) = (\pi/2) [q^{2} + k'^{2}-k^{2}]$. Therefore,
using the procedure described previously, the integral in Eq.(\ref{eq29})
is calculated and expressed in terms of the $\mathcal{J}_{p}(z)$ functions
defined in Appendix C,
\begin{eqnarray}
&&P_{12}=\frac{\hbar D^{2}A^{2}E_{F}^{2}(k_{B}T)^{4}}{\pi^{2}\rho_{m}(\hbar v_{s})^{3}(\hbar v_{F})^{5}}\left[
\mathcal{J}_{4}(T/\Theta_{BG}) - \frac{v_{s}}{v_{F}}\left(\frac{T}{\Theta_{BG}} \right)\right.\nonumber\\
&&\left.\times\mathcal{J}_{5}(T/\Theta_{BG})+
\frac{8 \pi^{2}}{3}\left(\frac{T}{\Theta_{BG}} \right)^{2}\left\{
\mathcal{J}_{4}(T/\Theta_{BG})+ \frac{3}{2}
\right.\right.\nonumber\\
&&\left.\left.\times
\left(\frac{v_{s}}{v_{F}} \right)^{2}\mathcal{J}_{4}(T/\Theta_{BG}) + \frac{1}{4\pi^{2}}\mathcal{J}_{6}(T/\Theta_{BG})
\right\}
+\frac{2}{3}
\right.\\
&&\left.\times
\left(\frac{v_{s}}{v_{F}} \right)^{3}\left(\frac{T}{\Theta_{BG}} \right)^{3}\left\{\mathcal{J}_{7}(T/\Theta_{BG}) - 2\pi^{2}\mathcal{J}_{5}(T/\Theta_{BG}) \right\}
\right]\nonumber
\label{eq31}
\end{eqnarray}

\section{Electrical Resistivity}
As discussed in Section III, the electrical resistivity is
obtained by setting $\nabla T = 0$ in Eq. (\ref{eq10}), to obtain 
$\rho = L_{EE}^{-1}$. By direct substitution of the coefficients
and currents obtained in Section III, we obtain after Eqs.(\ref{eq19}) that the leading 
term which defines the
electrical resistivity due to electron-phonon scattering processes is
\begin{eqnarray}
&&\rho_{e-ph}(T) = \rho_{0}\left[\left(\frac{T}{\Theta_{BG}}\right)^4{\mathcal{J}}_4(\Theta_{BG}/T) + \frac{4\pi^{2}}{3}\left(\frac{v_{s}}{v_{F}} \right)^{2}\right.\nonumber\\
&&\left.\times\left(\frac{T}{\Theta_{BG}}\right)^6\left\{{\mathcal{J}}_4(\Theta_{BG}/T)
+\frac{3}{2\pi^{2}}{\mathcal{J}}_6(\Theta_{BG}/T)\right\}\right]
\label{eq33}
\end{eqnarray}
Here, $\rho_{0} = 8\,D^{2}k_{F}/(e^{2}\rho_{m}v_{s}v_{F}^{2})$
is a coefficient with dimensions of a two-dimensional resistivity ($\Omega$),
and $\Theta_{BG}=2(v_{s}/v_{F})E_{F}/k_{B}$ is the Bloch-Gr\"uneisen temperature.
The functions ${\mathcal{J}}_{p}(\Theta_{BG}/T)$ and their asymptotic properties
are defined in Appendix C.
Eq. (\ref{eq33}) depends on the dimensionless parameter $T/\Theta_{BG}$,
a feature previously obtained\cite{Hwang.08} within the relaxation
time approximation. The functions ${\mathcal{J}}_p(\Theta_{BG}/T)$, defined
in Appendix C, have the property ${\mathcal{J}}_p(\infty) = p!\zeta(p)$, with $\zeta(p)$ the Riemann zeta function. Therefore,
the low-temperature ($T\ll \Theta_{BG}$) behavior of the electron-phonon
contribution to the electrical resistivity is
\begin{eqnarray}
\rho_{e-ph}(T) &&= \rho_{0}\,4!\zeta(4)(T/\Theta_{BG})^{4} + o(T/\Theta_{BG})^{6}
\label{eq34}
\end{eqnarray}
At high temperature ($T \gg \Theta_{BG}$), from Eq.(\ref{eq33}) we find the asymptotic
limit 
$\rho_{e-ph}(T) = \{1/3  - 1/10 \ldots \} \rho_{0}\, T/\Theta_{BG} =
(\pi/8)\rho_{0}\,T/\Theta_{BG}$.
This behavior is in agreement with experiments \cite{Efetov.010} and earlier theoretical results based on the relaxation time approximation \cite{Hwang.08}, thus
supporting the application of the variational method in this case. In Fig.(\ref{fig1}) we compare the theoretical prediction with the experimental data
reported in Ref. (\cite{Efetov.010}). The experimental data are fitted to
the expression
\begin{eqnarray}
\rho(T) = \rho_{(T=0)} + \rho_{e-ph}(T),
\label{eq35}
\end{eqnarray}
with $\rho_{e-ph}(T)$ defined by our theory Eq.(\ref{eq33}), and $\rho_{(T=0)}$ being the value of the resistivity
at zero temperature. This parameter can be interpreted as the (temperature independent) contribution of impurity scattering to the total electrical
resistivity. The impurity contribution to the electrical resistivity
has been discussed \cite{Stauber.07,Peres.07}, and it is a sample-dependent property, since
it is proportional to the surface concentration of impurities $n_{imp}$. It has been shown \cite{Stauber.07,Peres.07} that, for a short range scattering potential $V_{scatt}(\mathbf{r}) = V_{0}\delta(\mathbf{r})$, the in-plane resistivity is $\rho_{scatt} \propto n_{imp} V_{0}^{2}$, whereas for a long range Coulomb potential \cite{Stauber.07} $\rho_{scatt} \propto n_{imp} n^{-1}$, this last case showing an inverse dependency on the carrier concentration $n$
as well. Therefore, it is expected that the zero temperature
resistivity exhibits a dependence on the carrier concentration
of the form $\rho_{(T=0)} = a + b n^{-1}$, with the value of the coefficients $a$ and $b$ depending on details of the sample, particularly on the concentration and distribution of impurities and defects. We verify that this equation is in excellent agreement with the data reported in Ref. (\cite{Efetov.010}), as shown in the inset of Fig. (\ref{fig1}). We extracted values $a = 14.276$ ($\Omega$) and $b = 266.76\times 10^{17}$ ($\Omega\, \rm{m^{2}}$), with linear regression coefficient $r=0.999$. 
 
As seen in Fig. (\ref{fig1}), Eq.(\ref{eq35}) provides
an excellent fit to the experimental data. The temperature-dependent electron-phonon contribution to the electrical resistivity is fitted with just two free parameters: The energy parameter $D = 23.5 \pm 0.5$ eV in the deformation potential, and the speed of sound for longitudinal
phonons in graphene $v_{s} = 24.0 \pm 0.6$ Km/s. It is remarkable that
both values are in excellent agreement with independent estimations
reported in the literature, particularly the speed of sound \cite{Munoz.010, Saito.98}. It is also relevant to notice that the ratio $(v_{s}/v_{F})^{2}\sim 10^{-4}$ is negligibly small, and hence terms proportional
to this factor can in practice be neglected in the analytical expressions
obtained for the electrical resistivity, as well as in other transport coefficients
discussed along this work. This also explains the quantitative success 
when applying the relaxation time
approximation to electron-phonon scattering as reported in previous theoretical work \cite{Hwang.08}, even though this scattering mechanism is not strictly quasi-elastic.

\section{Thermal Resistivity}

The thermal conductivity is obtained from the general expression Eq. (\ref{eq10}) by setting $\mathbf{J} = 0$, which as discussed in Section III
leads to the expression $\kappa = L_{TE}L_{EE}^{-1}L_{ET} - L_{TT}$.
By direct substitution of the coefficients and currents calculated in
Section III, and neglecting terms proportional to $(v_{s}/v_{F})^{2}\sim 10^{-4}$, we obtain that the leading contribution to the thermal resistivity due to electron-phonon scattering is given by
\begin{eqnarray}
&&[\kappa_{e-ph}(T)]^{-1}=
\frac{\rho_{0}}{{\mathcal{L}}_{0}T}\left\{\left[\left(T/\Theta_{BG}\right)^4
+\frac{3}{4\pi^{2}}(T/\Theta_{BG})^{2}\right]\right.\nonumber\\
&&\times\left.{\mathcal{J}}_4(\Theta_{BG}/T)
-\frac{1}{2\pi^{2}}\left(T/\Theta_{BG}\right)^{4}{\mathcal{J}}_6(\Theta_{BG}/T)
\right\}
\label{eq37}
\end{eqnarray}
Here, ${\mathcal{L}}_{0} = \pi^{2}k_{B}^{2}/(3\,e^{2})$ is the Lorenz number
for the free electron gas.
At very low temperatures ($T\ll \Theta_{BG}$), Eq. (\ref{eq37}) predicts
the asymptotic limit
\begin{eqnarray}
[\kappa_{e-ph}(T)]^{-1} =
\frac{\rho_{0}}{{\mathcal{L}}_{0}\Theta_{BG}}\frac{3}{4\pi^{2}}4!\zeta(4)\frac{T}{\Theta_{BG}}+o\left(\frac{T}{\Theta_{BG}}\right)^{3}
\label{eq38}
\end{eqnarray}
 It is remarkable that, according to Eq. (\ref{eq38}), the
phonon-limited thermal resistivity is linear at very low temperatures, in contrast
with the typical $\sim T^{2}$ behavior theoretically predicted and experimentally
observed in normal, three-dimensional metals \cite{Ziman.60,Madelung.78}. 
The thermal resistivity contribution due to elastic scattering with Coulomb impurities has been calculated\cite{Peres.07}, and corresponds to $[\kappa_{scatt}]^{-1} = 3hT^{-1}u_{0}^{2}/(2\pi^{2}k_{B}^{2}E_{F}^{2})$,
with $u_{0}^{2}=n_{imp}Ze^{2}/(16\epsilon^{2}\epsilon_{0}^{2})$. 
The total thermal resistivity can be estimated from the expression $[\kappa(T)]^{-1} \sim [\kappa_{e-ph}(T)]^{-1} + [\kappa_{scatt}(T)]^{-1}$.
In Fig.(\ref{fig2}) inset,
we compare the value of the total three-dimensional thermal conductivity at room temperature, normalized
by a nominal packing "thickness" of $3.4$ \AA \ for the graphene layer. For numerical evaluation, we assumed $Z=1$ for the impurities, and
from the experimental values of the zero temperature electrical
resistivity, we extracted an average impurity concentration $n_{imp} = 1.3\times10^{15}$ $m^{-1}$. For the experimental system reported in
Ref.(\cite{Efetov.010}), we estimated a dielectric constant $\epsilon = 3.1$
representing the average between the $SiO_{2}$ substrate and the PEO polymer electrolyte. As seen in the inset of Fig.(\ref{fig2}), the total
normalized thermal conductivity at room temperature is on the 
order of $\kappa \sim 400$ $\rm{W} \rm{m}^{-1}$ $\rm{K}^{-1}$. We can compare
this with the thermal conductivity due to the phonon system, where the
experimental value\cite{Ghosh.08,Balandin.08}, in excellent agreement with a theory previously reported by the present author\cite{Munoz.010}, is
about\cite{Munoz.010,Ghosh.08,Balandin.08} $\tilde{\kappa}_{ph} = 4\,300$ $\rm{W} \rm{m}^{-1} \rm{K}^{-1}$ at room temperature. These results suggest that, in most of the temperature range,
the phonon contribution to the thermal conductivity in graphene dominates over the
electronic contribution, as the present author has pointed out elsewhere \cite{Munoz.010}.

\section{Thermopower}
As discussed in Section III, by setting $\mathbf{J}=0$ in Eq.(\ref{eq10}),
we obtain that the thermopower (Seebeck coefficient) $Q$ is given by
$Q = -L_{ET}/L_{EE}$. By direct substitution of
Eq.(\ref{eq19}), we obtain that the leading contribution
to the thermopower is given by the expression
\begin{widetext}
\begin{eqnarray}
Q(T) = -\frac{\pi^{2}}{3e}\frac{k_{B}^{2}T}{E_{F}}
\frac{\left[1 + \frac{4\pi^{2}}{3}\left(\frac{T}{\Theta_{BG}} \right)^{2} \right]\mathcal{J}_{4}(\Theta_{BG}/T) + \frac{v_{s}}{v_{F}}\left(\frac{T}{\Theta_{BG}} \right)\mathcal{J}_{5}(T/\Theta_{BG})-2\left(\frac{T}{\Theta_{BG}} \right)^{2}\mathcal{J}_{6}(\Theta_{BG}/T)}
{\left[1 + \frac{4\pi^{2}}{3}\left(\frac{T}{\Theta_{BG}} \right)^{2}  \right]\mathcal{J}_{4}(\Theta_{BG}/T) - \frac{2}{3}\left(\frac{T}{\Theta_{BG}} \right)^{2}\mathcal{J}_{6}(T/\Theta_{BG})}
\label{eq41}
\end{eqnarray}
\end{widetext}

\begin{figure}[tbp]
\centering
\epsfig{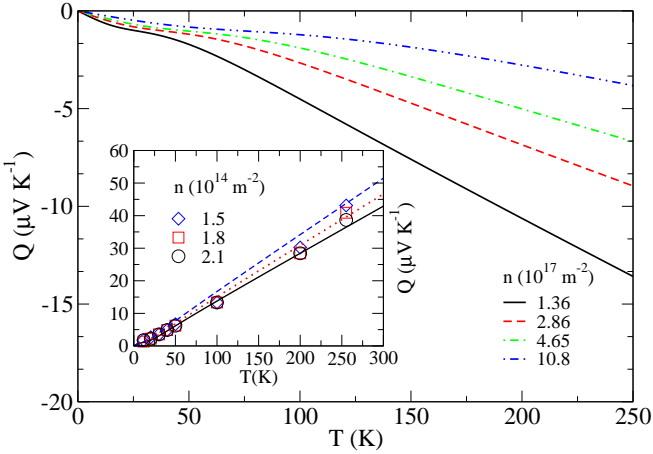}
\caption{(Color online) Thermopower at different electronic densities $n$, in units of $10^{17}$ ${\rm{m}}^{-2}$, calculated after Eq.(\ref{eq41}) with $v_{s} = 24$ Km/s as inferred from experimental data in Ref.(\cite{Efetov.010}). Inset: Experimental data (symbols) for hole thermopower in Ref.(\cite{Wei.09}), compared
with Eq.(\ref{eq41}) at the corresponding hole densities (no fitting parameters besides the value of $v_{s} = 24$ Km/s are used).
\label{fig3}
}
\end{figure}
Therefore, we find that the thermopower at very low temperatures ($T\ll \Theta_{BG}$) becomes
\begin{eqnarray}
Q(T) \sim -\frac{\pi^{2}}{3e}\frac{k_{B}^{2}T}{E_{F}}\left[1 
+\frac{v_{s}}{v_{F}}\frac{5!\zeta(5)}{4!\zeta(4)}\left(\frac{T}{\Theta_{BG}}
\right)
\right] + o(T^{3})
\label{eq42}
\end{eqnarray}
This result is in agreement with experimental data reported in the literature \cite{Wei.09}. In particular, it has been observed \cite{Wei.09} that
the thermopower in graphene depends on the carrier density as $n^{-1/2}$,
and shows a linear in $T$ dependence at very low temperatures.  
Since the Fermi level is $E_{F} = \hbar v_{F}\sqrt{\pi n}$, it is clear
that Eq.(\ref{eq41}) obtained from our theory correctly reproduces
this feature. The contribution to the total thermopower due to phonon-drag effects has been discussed in Ref.(\cite{Kubakaddi.09}), where it is shown
that it displays a $T^{3}$ dependence at low temperatures. Clearly then, the phonon drag is a negligible contribution to the total thermopower as compared to the diffusion component calculated in this work, which
in agreement with the experimental data \cite{Wei.09} reproduces the correct linear in $T$ dependence at very low temperatures. 

\section{Conclusions}

We presented a semi-classical theory to calculate the phonon-limited transport coefficients in extrinsic graphene at high carrier densities. Our theory is based on a variational solution of the Boltzmann equation, and provides explicit analytical expressions for the various transport coefficients, such as electrical resistivity, thermal resistivity and
thermopower. We showed that our analytical results are in excellent agreement with experimental data arising from two independent groups,
particularly concerning the electrical resistivity \cite{Efetov.010}
and thermopower \cite{Wei.09}.

Our theoretical results suggest that, in principle, it is possible to control not only the electrical resistivity, but also the thermal resistivity and thermopower by
controlling the extrinsic carrier concentration in graphene.

\section{Acknowledgements}
The author wish to thank FONDECYT grant $\rm{N^{o}}$ 11100064
for financial support.

\appendix

\section{Evaluation of angular integrals}

We consider the elementary integrals over angular orientations of the
wave-vectors used in the main text. Let us define the unit vector $\hat{u}=(\cos(\alpha),\sin(\alpha))$, with $\alpha$ an arbitrary (but fixed) direction along the thermal gradient or the applied electric field.
As in the main text, we define the scattering angle $\theta$ by $\hat{k}\cdot\hat{k}'=\cos(\theta)$, with $\hat{k}$, $\hat{k}'$ the directions of the initial and final wave-vector after the electron-phonon scattering event takes place. Then, we write $\hat{k}'=(\cos(\varphi'),\sin(\varphi'))$ and $\hat{k}=(\cos(\varphi'+\theta),\sin(\varphi'+\theta))$, respectively. Using this notation, we consider the following integrals:
\begin{eqnarray}
&&\int_{-\pi}^{+\pi}d\varphi' [\hat{k}\cdot \hat{u}]^{2} \nonumber\\
&&= \int_{-\pi}^{+\pi}d\varphi'\left[\cos(\alpha)\cos(\varphi'+\theta)+
\sin(\alpha)\sin(\varphi'+\theta) \right]^{2}\nonumber\\
&&= \pi[(\cos(\alpha))^{2} + (\sin(\alpha))^{2}] = \pi
\end{eqnarray}
\begin{eqnarray}
&&\int_{-\pi}^{+\pi}d\varphi' [\hat{k}'\cdot \hat{u}]^{2}\nonumber\\
&&= \int_{-\pi}^{+\pi}d\varphi'\left[\cos(\alpha)\cos(\varphi')+
\sin(\alpha)\sin(\varphi') \right]^{2}\nonumber\\
&&= \pi[(\cos(\alpha))^{2} + (\sin(\alpha))^{2}] = \pi
\end{eqnarray}
\begin{eqnarray}
&&\int_{-\pi}^{+\pi}d\varphi' [\hat{k}\cdot \hat{u}][\hat{k}'\cdot \hat{u}] \nonumber\\
&&=\int_{-\pi}^{+\pi}d\varphi'\left[\cos(\alpha)\cos(\varphi'+\theta)+
\sin(\alpha)\sin(\varphi'+\theta) \right]\nonumber\\
&&\times\left[\cos(\alpha)\cos(\varphi')+
\sin(\alpha)\sin(\varphi')\right]\nonumber\\
&&=\int_{-\pi}^{+\pi}d\varphi'\left[(\cos(\alpha))^{2}\cos(\varphi'+\theta)\cos(\varphi')\right.\nonumber\\
&&\left.+(\sin(\alpha))^{2}\sin(\varphi'+\theta)\sin(\varphi') + \sin(\alpha)\cos(\alpha)\right.\nonumber\\
&&\left.\times
\sin(2\varphi'+\theta) \right]
=\pi\cos(\theta)
\end{eqnarray}

Direct application of these basic identities leads to the following results
\begin{eqnarray}
&&\int_{-\pi}^{+\pi}d\varphi'[\mathbf{k}\cdot \hat{u}][\mathbf{k}'\cdot \hat{u}]=k k'\int_{-\pi}^{+\pi}d\varphi'[\hat{k}\cdot \hat{u}][\hat{k}'\cdot \hat{u}]\nonumber\\
&&=\pi k k'\cos(\theta)
\end{eqnarray}
\begin{eqnarray}
&&\int_{-\pi}^{+\pi}d\varphi'[\mathbf{q}\cdot \hat{u}]^{2}=\int_{-\pi}^{+\pi}d\varphi'[(\mathbf{k}'-\mathbf{k})\cdot \hat{u}]^{2}\nonumber\\
&&=\pi(k^{2} + k'^{2} - k k'\cos(\theta) )=\pi(\mathbf{k}'-\mathbf{k})^{2}\nonumber\\
&&=\pi q^{2}
\end{eqnarray}
\begin{eqnarray}
&&\int_{-\pi}^{+\pi}d\varphi'[\mathbf{q}\cdot\hat{u}][\mathbf{k}'\cdot\hat{u}]=\int_{-\pi}^{+\pi}d\varphi'[(\mathbf{k}'-\mathbf{k})\cdot \hat{u}][\mathbf{k}'\cdot\hat{u}]\nonumber\\
&&=\pi(k'^{2} - 2k k'\cos(\theta))=\pi\left(\frac{q^{2}}{2} +\frac{k'^{2}- k^{2}}{2}\right)
\end{eqnarray}

\section{Evaluation of energy integrals at low temperature}

In several calculations throughout this work,
after the change of variables $x = (\epsilon - E_{F})/(k_{B}T)$ and
$z=\hbar\omega/(k_{B}T)$, we use the identity
\begin{eqnarray}
\int_{-\infty}^{+\infty}\frac{F(x)}{\left(1 + e^{x} \right)\left(1 + e^{-x-z} \right)}dx = \left(1 - e^{-z} \right)^{-1}\nonumber\\
\times\int_{-\infty}^{+\infty}\left[ G(x) - G(x - z) \right]\left(-\frac{\partial f^{0}}{\partial x} \right)dx
\end{eqnarray}
with $G(x) = \int_{0}^{x}F(x')d x'$. This identity is straightforward to prove using integration by parts, and noting that $f^{0}(x) - f^{0}(x + z) =
(1 - e^{-z})(1 + e^{x})^{-1}(1 + e^{-x-z})^{-1}$.

The right-hand side of the equation can be further evaluated using the Sommerfeld expansion for the Fermi-Dirac distribution function at low temperatures, leading to the formula
\begin{eqnarray}
&&\int_{-\infty}^{+\infty}\frac{F(x)}{\left(1 + e^{x} \right)\left(1 + e^{-x-z} \right)}dx\nonumber\\
%= (1 - e^{-z})^{-1}\left\{G(0) - G(-z) + \frac{\pi^{2}}{6}
%[G^{''}(0) - G^{''}(-z)] \right\}\nonumber\\
&&=(1 - e^{-z})^{-1}\left\{-\int_{0}^{-z}F(x)d x + \frac{\pi^{2}}{6}
[F^{'}(0) - F^{'}(-z)] \right\}\nonumber\\
\end{eqnarray}
which is valid up to $o((k_{B}T/E_{F})^{4})$.

\section{Definition and properties of the functions $\mathcal{J}_{p}(z)$}

We have defined the functions
\begin{eqnarray}
{\mathcal{J}}_{p}(z) =
\int_{0}^{z}\frac{x^{p}\sqrt{1-(x/z)^{2}}}{(1-e^{-x})(e^{x}-1)}dx
\end{eqnarray}
The asymptotic limit of these functions, for $p>0$ an integer, is given by
${\mathcal{J}}_{p}(\infty) = p!\zeta(p)$,
with $\zeta(p)$ the Riemann Zeta function. In particular, for $p=4$ and $6$ it has the values $\zeta(4) = \pi^{4}/90$ and $\zeta(6) = \pi^{6}/945$.

\end{document}